# AI Use Conditions and Perspective Diversity in Ethical Decision-Making: A Pilot Study of Human Reasoning Processes

Byeongmu Choi

Independent Researcher
Republic of Korea

Corresponding Author:
Byeongmu Choi
Email: qudan5505@naver.com

Abstract

Generative artificial intelligence (AI) is increasingly used to support human decision-making. While prior research has primarily focused on outcomes such as productivity, performance, and decision quality, less attention has been paid to how AI may influence the reasoning processes that precede final judgments. This pilot study explored whether different AI-use conditions were associated with differences in reasoning breadth during ethical decision-making.

Twenty-nine participants completed an ethical dilemma involving whether a company should disclose a known product defect to customers. Participants were recruited separately into one of three conditions: AI-Prohibited (n = 10), AI-Optional (n = 10), or AI-Mandatory (n = 9). Responses were evaluated by three independent blind coders using a predefined coding manual. Two exploratory measures were employed: the Counterargument Diversity Score (CDS) and the Perspective Diversity Index (PDI).

Across all conditions, participants generally converged on similar ethical conclusions, most commonly favoring disclosure and customer protection. However, differences emerged in the reasoning processes used to reach those conclusions. Participants in the AI-Mandatory condition considered a broader range of perspectives, including legal, regulatory, organizational, technical, and ethical viewpoints. A statistically significant overall difference in Perspective Diversity Index scores was observed across the three conditions, with the AI-Mandatory group showing substantially higher mean scores than the AI-Prohibited and AI-Optional groups. In contrast, differences in Counterargument Diversity Scores were not statistically significant.

These findings suggest that generative AI may not necessarily alter final ethical judgments, but may be associated with broader exploration of perspectives prior to reaching those judgments. Although the results should be interpreted cautiously given the small sample size, they highlight the importance of examining how AI influences reasoning processes rather than focusing exclusively on decision outcomes.

## 1. Introduction

Generative artificial intelligence (AI) has rapidly become part of everyday decision-making and problem-solving activities. Large language models (LLMs), such as ChatGPT, are increasingly used not only for information retrieval but also for idea generation, analytical support, and reflective reasoning. As a result, generative AI is beginning to function as a cognitive support tool that may influence how individuals evaluate information and formulate judgments.

Recent studies have demonstrated that generative AI can improve task performance across a variety of domains (Noy & Zhang, 2023; Dell'Acqua et al., 2026). For example, Noy and Zhang (2023) reported substantial productivity gains among knowledge workers using generative AI, while Dell’Acqua et al. (2026) found that AI assistance improved the quality of consultant performance on selected professional tasks. Collectively, these studies suggest that AI systems can meaningfully influence human work and decision-making outcomes.

However, most existing research has focused on outcomes, including productivity, accuracy, performance quality, and task completion. Comparatively less attention has been devoted to understanding whether AI influences the reasoning processes that precede those outcomes. This distinction is important because individuals may arrive at similar conclusions while considering very different sets of perspectives, stakeholders, risks, and counterarguments along the way.

From a Human-Centered AI perspective (Shneiderman, 2022), the key question is not only whether AI changes decisions, but also whether it changes how people think about decisions. Human judgment is often shaped by the range of perspectives considered before a conclusion is reached. Prior research in judgment and decision-making has suggested that actively considering alternative viewpoints and opposing arguments can reduce cognitive biases and promote more balanced reasoning (Lord et al., 1984). Yet relatively little is known about whether generative AI affects this process of perspective exploration.

More broadly, research on judgment and decision-making has shown that individuals frequently rely on simplified heuristics when evaluating complex situations (Tversky & Kahneman, 1974). Likewise, research on automation has emphasized that technological systems can influence how people process information, allocate attention, and form judgments (Lee & See, 2004; Parasuraman & Riley, 1997). Despite these insights, relatively little research has examined whether generative AI influences the breadth of reasoning that precedes final decisions. Understanding this process may be particularly important because individuals can reach similar conclusions while considering substantially different sets of perspectives, stakeholders, risks, and counterarguments.

The present study focuses on this gap. Rather than examining whether AI changes final ethical judgments, we investigate whether different AI-use conditions are associated with differences in the breadth of reasoning that precedes those judgments. Specifically, we examine the diversity of perspectives and counterarguments considered by participants while responding to an ethical dilemma.

Importantly, the concept of perspective diversity in this study does not imply decision quality or correctness. Individuals may reach identical conclusions while relying on different interpretive frames and considering different stakeholders. Perspective diversity is therefore defined as the extent to which participants examined the dilemma through multiple independent perspectives, including customer, employee, shareholder, legal, regulatory, ethical, organizational, technical, and societal viewpoints.

To assess this construct, we developed an exploratory measure called the Perspective Diversity Index (PDI). The PDI was designed to capture the number of distinct perspective categories considered during reasoning rather than to evaluate the quality of the final decision. We also assessed Counterargument Diversity Score (CDS), which measures the diversity of opposing arguments participants considered against their own position.

Participants completed the same ethical dilemma under one of three conditions: AI-Prohibited, AI-Optional, or AI-Mandatory. Responses were evaluated using a predefined coding manual and independently coded by three blind raters.

The study addressed the following research questions:

RQ1. Are different AI-use conditions associated with differences in counterargument diversity during ethical decision-making?

RQ2. Are different AI-use conditions associated with differences in perspective diversity during ethical decision-making?

As a pilot study, the goal is not to establish causal claims regarding the effects of generative AI, but to explore whether AI-use conditions are associated with differences in reasoning processes that may warrant further investigation in larger and more rigorous studies.

## 2. Method

### 2.1 Study Design

This pilot study employed a between-subjects design to explore whether different AI-use conditions were associated with differences in reasoning processes during ethical decision-making.

Participants were recruited separately for one of three conditions:

- AI-Prohibited
- AI-Optional
- AI-Mandatory

All participants completed the same ethical dilemma task, received the same instructions regarding perspective-taking, and answered the same set of open-ended questions. The only difference between conditions concerned the use of generative AI tools.

The primary objective was not to compare final decisions, but rather to examine differences in the breadth of reasoning used to reach those decisions.

**2.2 Participants**

A total of 30 participants were recruited through Prolific.

Eligibility criteria required participants to have a Prolific approval rate of at least 95% and between 50 and 5,000 prior submissions. Participants were recruited from English-speaking countries, including the United States, the United Kingdom, Canada, Australia, and Ireland.

All conditions were advertised as approximately 12-minute studies and offered compensation of £3.

One participant recruited for the AI-Mandatory condition reported not using AI and failed to provide the required evidence of AI use. Consistent with the predefined study protocol, this participant received compensation but was excluded from the final analysis.

The final sample consisted of 29 participants:

AI-Prohibited (n = 10),
AI-Optional (n = 10),
AI-Mandatory (n = 9).

Because the study was designed as an exploratory pilot study, demographic information such as age, gender, and occupation was not collected.

Prior to participation, all participants provided informed consent and were informed that participation was voluntary, anonymous, and could be discontinued at any time. Participants confirmed that they were at least 18 years old, voluntarily agreed to participate, and understood that their responses could be used anonymously for research purposes. No names, personal email addresses, or other direct identifiers were collected. Prolific used platform-generated pseudonymous participant identifiers for study administration and compensation, and these identifiers were not retained in the dataset used for coding and statistical analysis.

This study was conducted as an independent research project and was not reviewed by an institutional review board or research ethics committee. At the time the study was conducted, the author was not affiliated with an institution that provided access to a formal ethics review process. No formal exemption determination was obtained.

The study involved a hypothetical organizational dilemma, did not involve deception or vulnerable populations, and did not request sensitive personal information.

2.3 Ethical Dilemma Task

All participants completed the same ethical decision-making task.

The scenario described a medium-sized company that was aware of a product defect but had chosen not to fully disclose the issue to customers. Although the issue had not been formally determined to be illegal, some customers may already have experienced harm.

The company was reluctant to disclose the problem because of concerns regarding reputational damage, stock price decline, potential lawsuits, refund costs, and possible workforce reductions.

Participants were asked to explain how the company should respond and justify their reasoning.

The dilemma was intentionally selected because most participants were expected to converge on broadly similar ethical conclusions. This design choice allowed the study to focus on differences in reasoning processes rather than differences in final judgments.

Participants responded to five open-ended questions concerning:

(1) recommended actions,

(2) relevant stakeholders,

(3) counterarguments,

(4) risks, and

(5) alternative perspectives.

## 2.4 Experimental Conditions

All participants were encouraged to consider multiple perspectives before providing their final responses.

Specifically, participants were instructed to consider:

- different stakeholders,
- possible counterarguments,
- potential risks, and
- alternative perspectives.

Participants were also informed that there were no objectively correct answers and that the study focused on their reasoning processes.

### AI-Prohibited Condition

Participants were instructed not to use generative AI tools, search engines, or assistance from other individuals while completing the task.

### AI-Optional Condition

Participants were informed that they could use generative AI tools such as ChatGPT, Claude, or Gemini if they wished.

However, they were instructed not to copy AI-generated responses directly and instead to write their final responses in their own words.

Following task completion, participants were invited to describe how they had used AI. This question was optional.

Among those who responded, six participants reported not using AI, one participant reported using AI as a discussion partner, and three participants did not answer the question.

### AI-Mandatory Condition

Participants were informed that using a generative AI tool was required.

They were instructed to interact with AI before submitting their final responses.

To verify compliance, participants were required to provide a brief description of their AI use and submit the first AI interaction generated during task completion.

Submission of the first AI interaction was mandatory, whereas submission of the full AI conversation history was optional.

### 2.5 Coding and Measures

Participant responses were evaluated using the predefined coding manual presented in Appendix A. Three independent coders who were blind to participants' experimental condition evaluated all responses. Coders were instructed to score only explicitly stated content, avoid inference, and adopt a conservative interpretation when uncertainty arose. The coding procedure was intentionally conservative. To avoid inflating scores merely because a response was longer or more procedurally detailed, coders counted only distinct, explicitly stated perspective or counterargument categories. Repeated mentions of the same category were counted only once, and actions, procedures, recommendations, and outcome descriptions were not counted as perspective categories. This reduced the likelihood that AI-assisted responses would receive higher PDI scores simply because they were more verbose or solution-oriented. Two exploratory measures were calculated: Counterargument Diversity Score (CDS) and Perspective Diversity Index (PDI). Inter-rater reliability was assessed using ICC(2,k).

## 3. Results

### 3.1 Inter-Rater Reliability

Three independent coders evaluated all participant responses using the predefined coding manual.

To assess coding consistency, inter-rater reliability was calculated using a two-way random-effects intraclass correlation coefficient, ICC(2,k) (Shrout & Fleiss, 1979).

Table 1 presents the reliability estimates for both outcome measures.

Table 1. Inter-Rater Reliability Across Three Independent Coders

| Measure | ICC(2,k) |
|---|---|
| CDS | 0.868 |
| PDI | 0.817 |

The ICC values indicated good agreement among coders for both measures. According to commonly used reliability guidelines, both coefficients suggest that the coding manual was applied with a satisfactory degree of consistency across raters.

3.2 AI Use Compliance

Participants in the AI-Optional condition were invited to describe how they used AI during the task.

Among the ten participants in this condition, seven responded to the optional question. Six reported that they did not use AI, one reported using AI as a discussion partner, and three did not provide a response.

In the AI-Mandatory condition, participants were required to use a generative AI tool and submit evidence of AI use. One participant failed to comply with this requirement and was excluded from the final analysis. The remaining nine participants provided the required evidence and were included in the study.

3.3 Counterargument Diversity Score (CDS)

Mean CDS values for each condition are presented in Table 2.

Table 2. Counterargument Diversity Score (CDS) by Condition

| Condition | n | Mean | SD |
|---|---|---|---|
| AI-Prohibited | 10 | 1.47 | 1.34 |
| AI-Optional | 10 | 1.53 | 0.91 |
| AI-Mandatory | 9 | 1.96 | 1.30 |

Participants in the AI-Mandatory condition achieved the highest average CDS score. However, differences among conditions were relatively modest.

A one-way analysis of variance (ANOVA) revealed no statistically significant difference among the three conditions, $F(2,26) = 0.48$, $p = .626$, $\eta^2 = .035$.

These results suggest that AI-use condition was not strongly associated with differences in the diversity of counterarguments considered by participants.

### 3.4 Perspective Diversity Index (PDI)

Mean PDI values for each condition are presented in Table 3.

Table 3. Perspective Diversity Index (PDI) by Condition

| Condition | n | Mean | SD |
|---|---|---|---|
| AI-Prohibited | 10 | 0.87 | 0.61 |
| AI-Optional | 10 | 0.87 | 0.67 |
| AI-Mandatory | 9 | 2.30 | 1.72 |

The AI-Prohibited and AI-Optional groups produced identical mean PDI scores, whereas the AI-Mandatory group demonstrated substantially higher perspective diversity.

A one-way ANOVA revealed a statistically significant effect of condition on PDI, $F(2,26) = 5.31$, $p = .012$, $\eta^2 = .290$.

To provide an estimate of practical significance, Cohen's d was calculated for pairwise comparisons. The difference between the AI-Mandatory and AI-Prohibited conditions yielded a large effect size ($d = 1.13$). A similarly large effect was observed between the AI-Mandatory and AI-Optional conditions ($d = 1.12$).

These findings indicate that participants required to interact with AI considered a broader range of perspectives than participants in the other two conditions.

### 3.5 Summary of Findings

Across all three conditions, participants generally reached similar ethical conclusions, most commonly favoring disclosure of the product issue and prioritizing customer protection.

Despite this convergence in final judgments, differences emerged in the reasoning processes used to arrive at those conclusions.

No statistically significant differences were observed for Counterargument Diversity Score (CDS). In contrast, Perspective Diversity Index (PDI) differed significantly across conditions, with the AI-Mandatory group achieving substantially higher scores than both the AI-Prohibited and AI-Optional groups.

Taken together, these results suggest that AI-use conditions may be more strongly associated with the diversity of perspectives considered during reasoning than with the diversity of counterarguments generated against one's own position.

## 4. Discussion

The present pilot study explored whether different AI-use conditions were associated with differences in reasoning processes during ethical decision-making.

A notable finding was that participants across all three conditions generally arrived at similar ethical conclusions. Most participants supported disclosing the product issue to customers and emphasized customer protection as a primary consideration. In other words, the final judgments produced by participants were broadly similar regardless of AI-use condition.

However, differences emerged in the reasoning processes used to reach those judgments.

Participants in the AI-Mandatory condition demonstrated substantially higher Perspective Diversity Index (PDI) scores than participants in either the AI-Prohibited or AI-Optional conditions. In contrast, no statistically significant differences were observed for Counterargument Diversity Score (CDS).

It is important to emphasize that the PDI was designed to capture the diversity of perspectives considered during reasoning, rather than the depth, quality, or correctness of those perspectives. Therefore, higher PDI scores should be interpreted as indicating broader perspective exploration, not necessarily better reasoning or superior ethical judgment.

Notably, participants in the AI-Mandatory condition often referenced distinct interpretive frames, including legal, ethical, organizational, regulatory, and technical perspectives. This pattern suggests that higher PDI scores reflected more than simple category listing and may indicate broader exploration of alternative ways of interpreting the dilemma.

These findings suggest that the influence of generative AI may not primarily manifest through changing final decisions. Instead, AI interaction may be associated with the range of perspectives considered before a decision is reached.

This distinction may be important for understanding the role of AI in human decision-making. Individuals can arrive at similar conclusions while relying on very different reasoning processes. Two people may both support disclosure of a product issue, for example, yet one may primarily consider customer welfare whereas another may additionally consider legal obligations, regulatory expectations, organizational responsibility, technical causes, and long-term public trust.

The results of the present study suggest that AI-supported reasoning may involve a broader set of interpretive frames even when final conclusions remain unchanged.

One possible interpretation of these findings is that AI interaction influenced how participants explored the problem rather than the conclusions they ultimately reached. Although participants frequently converged on similar ethical judgments, those in the AI-Mandatory condition appeared to examine the dilemma through a broader range of interpretive frames. From this perspective, the findings may suggest a broader exploration of the problem space during reasoning rather than a change in the final decision itself.

Interestingly, the strongest differences emerged in perspective diversity rather than counterargument diversity. Participants in the AI-Mandatory condition did not necessarily generate substantially more opposing arguments against their own positions. Instead, they appeared more likely to examine the dilemma through multiple perspectives.

One possible interpretation is that generative AI may function less as a source of opposition and more as a source of reframing. Rather than encouraging users to reject their existing conclusions, AI systems may expose users to additional ways of interpreting the same problem. In the present study, participants in the AI-Mandatory condition more frequently referenced legal, regulatory, organizational, ethical, technical, and societal considerations. This pattern is consistent with the possibility that AI interaction can introduce additional interpretive frames into the reasoning process.

Another noteworthy observation concerns the AI-Optional condition. Participants in this condition produced results that were highly similar to those of the AI-Prohibited group. Self-reported responses indicated that most participants who answered the follow-up question did not actually use AI during task completion.

Because most participants in the AI-Optional condition either reported not using AI or did not provide information regarding AI use, this condition may be interpreted as reflecting AI availability rather than active AI engagement. Consequently, the observed pattern suggests that actual interaction with AI, rather than mere access to AI tools, may be more closely associated with broader perspective exploration.

Although the sample size was small and AI use was not directly monitored in this condition, the pattern suggests that merely allowing AI access may not be sufficient to influence reasoning processes. Instead, actual interaction with AI may be a more important factor than availability alone.

The findings should be interpreted cautiously. This study was designed as an exploratory pilot investigation rather than a confirmatory experiment. The sample size was small, only a single ethical dilemma was employed, and demographic variables were not collected. Furthermore, the low rate of reported AI use within the AI-Optional condition limits conclusions regarding voluntary AI adoption.

Despite these limitations, the study contributes to an emerging area of research examining how generative AI influences human reasoning rather than simply evaluating decision outcomes. Much of the existing literature has focused on productivity, task performance, or decision quality. The present findings suggest that reasoning processes themselves may represent an additional and potentially important area of investigation.

From a Human-Centered AI perspective, the findings are consistent with the possibility that generative AI may serve as a cognitive support tool that helps users explore multiple interpretive frames while preserving human responsibility for final judgments. Whether this pattern generalizes across other domains remains an open empirical question.

Future research should examine larger samples, multiple dilemma types, and additional indicators of reasoning breadth. Measures related to stakeholder consideration, risk exploration, alternative generation, and trade-off evaluation may help clarify whether the observed pattern extends beyond perspective diversity alone.

If future studies consistently find that AI-use conditions are associated with broader reasoning processes across different contexts, generative AI may be understood not only as a tool for improving outcomes but also as a tool that shapes how individuals approach complex decisions.

## 5. Limitations

Several limitations should be considered when interpreting the findings of this study.

First, this research was conducted as a pilot study with a relatively small sample size (N = 29). Although the observed differences in perspective diversity were substantial, the study was not designed to provide definitive evidence regarding the effects of generative AI on reasoning processes. Replication with larger samples is necessary.

Second, the study employed only a single ethical dilemma scenario. The scenario was intentionally selected because it was expected to produce broadly similar final judgments, thereby allowing differences in reasoning processes to be examined under a common decision context rather than being confounded with large differences in participants' ultimate conclusions.

However, this design choice may also have constrained the range and variability of some reasoning processes. Because the dilemma presented a relatively strong ethical case for disclosure and customer protection, participants may have had limited scope to generate genuinely distinct counterarguments, alternatives, trade-offs, competing risk assessments, or conflicts among stakeholder interests. This restriction may have reduced the sensitivity of the task to detect condition-related differences in these dimensions.

In particular, the nonsignificant differences in Counterargument Diversity Score should not necessarily be interpreted as evidence that AI interaction had no relationship with counterargument generation. An alternative explanation is that the structure of the dilemma itself limited the number and diversity of plausible counterarguments available to participants. Future studies should therefore compare dilemmas that vary in moral clarity, balance between competing options, and complexity of trade-offs.

It remains unclear whether the observed pattern would generalize to other contexts, such as public policy decisions, medical decision-making, organizational strategy, or technology governance.

Third, demographic variables such as age, gender, education, and occupation were not collected. As a result, potential relationships between participant characteristics and reasoning patterns could not be examined.

Fourth, participants in the AI-Optional condition rarely reported using AI. Consequently, this condition may be more accurately interpreted as a voluntary-use condition rather than an active AI-use condition. Future studies should directly measure and verify AI engagement to better understand the effects of voluntary AI adoption.

Finally, the measures used in this study were exploratory. Although the coding manual demonstrated good inter-rater reliability, both the Perspective Diversity Index (PDI) and Counterargument Diversity Score (CDS) represent initial attempts to operationalize reasoning breadth. Additional validation of these measures is needed.

### 5.1 Future Research

Future studies should examine whether the observed association between AI use and perspective diversity replicates across larger samples and multiple decision-making domains, including healthcare, public policy, organizational management, and technology governance.

Future research should also compare different forms of AI engagement. A useful design would include an AI-Prohibited condition, a general AI-Mandatory condition, and an interactive AI condition in which participants are required to ask follow-up questions, reconsider initial assumptions, or explore alternative interpretations before providing a final response. This comparison could help distinguish the effects of merely using AI from those of engaging in a more iterative and reflective human-AI interaction process.

Multiple dilemmas should also be used. These scenarios may be designed to produce broadly similar final conclusions while still allowing variation in how participants qualify, justify, and integrate those conclusions. This would make it possible to examine conditional judgment, integration of competing considerations, decision-path sophistication, and overall reasoning coherence even when final decisions remain similar.

Beyond the Perspective Diversity Index and Counterargument Diversity Score, future studies may incorporate stakeholder consideration, risk exploration, alternative generation, trade-off recognition, decision confidence, perceived difficulty, and reliance on AI.

## 6. Conclusion

This pilot study examined whether different AI-use conditions were associated with differences in reasoning processes during ethical decision-making.

Across all conditions, participants generally reached similar ethical conclusions, most commonly supporting disclosure of the product issue and prioritizing customer protection. However, meaningful differences emerged in the breadth of perspectives considered during reasoning.

A statistically significant overall difference in perspective diversity was observed across the three conditions, with the AI-Mandatory group showing substantially higher mean scores than the AI-Prohibited and AI-Optional groups. In contrast, no statistically significant differences were observed for counterargument diversity.

These findings suggest that the influence of generative AI may not necessarily be reflected in changes to final decisions. Instead, AI interaction may be associated with broader exploration of perspectives before those decisions are reached.

Although preliminary, the results contribute to a growing body of research examining how generative AI shapes human reasoning processes. From a Human-Centered AI perspective, the findings are consistent with the possibility that AI can function as a cognitive support tool that helps individuals explore multiple interpretive frames while retaining responsibility for final judgments.

Understanding how AI influences reasoning processes may represent an important complement to existing research that focuses primarily on productivity, performance, and decision outcomes.

## Declaration of Generative AI Use

During the preparation of this manuscript, the author used ChatGPT (OpenAI) to assist with language refinement, manuscript organization, and improvement of academic writing clarity.

The author reviewed, revised, and approved all manuscript content and takes full responsibility for the accuracy, integrity, and interpretation of the work.

## Data Availability Statement

The anonymized data and coding materials supporting the findings of this study are available from the author upon reasonable request.

**Appendix A. Reasoning Breadth Coding Manual (Version 1.0)**

**Purpose**

This coding manual was developed to assess the breadth of reasoning demonstrated by participants during ethical decision-making tasks.

The objective is not to evaluate whether a participant reached the "correct" conclusion, but rather to measure how many independent perspectives and counterarguments were considered before reaching that conclusion.

Two primary variables were coded:

1. Counterargument Diversity Score (CDS)
2. Perspective Diversity Index (PDI)

---

**Counterargument Diversity Score (CDS)**

**Question**

"What arguments could be made against your position?"

**Definition**

CDS measures how many distinct categories of counterarguments a participant considered against their own position.

Only independent counterargument categories receive points.

Repeated references to the same category receive only one point.

---

**Coding Categories**

**Financial**

Examples:

- profit loss
- shareholder loss
- stock price decline

- lawsuits
- recall costs
- refunds

Score: +1

---

**Reputation**

Examples:

- reputation damage
- loss of trust
- brand harm

Score: +1

---

**Employee**

Examples:

- layoffs
- restructuring
- employee harm

Score: +1

---

**Evidence Uncertainty**

Examples:

- insufficient evidence
- not yet proven
- incomplete investigation

Score: +1

---

**Customer Panic**

Examples:

- unnecessary panic
- overreaction
- public alarm

Score: +1

---

**Competitive Disadvantage**

Examples:

- competitors gain advantage
- market disadvantage

Score: +1

---

**Organizational Survival**

Examples:

- corporate collapse
- business failure
- existential crisis

Score: +1

---

**Scoring Examples**

Example A

"The company may lose money and suffer reputation damage."

Financial = 1

Reputation = 1

Total CDS = 2

---

Example B

"The evidence is incomplete and disclosure could create unnecessary panic."

Evidence = 1

Customer Panic = 1

Total CDS = 2

---

**Perspective Diversity Index (PDI)**

**Question**

"Are there alternative ways to view this problem?"

**Definition**

PDI measures how many independent perspectives participants used to interpret the dilemma.

Only perspectives receive points.

Actions, solutions, recommendations, and outcomes do not receive points.

---

**Coding Categories**

**Customer Perspective**

Examples:

- customer rights
- customer safety
- customer protection

Score: +1

---

**Shareholder Perspective**

Examples:

- shareholder interests
- investor concerns

Score: +1

---

**Employee Perspective**

Examples:

- employee welfare
- workforce impact

Score: +1

---

**Management Perspective**

Examples:

- business interests
- management priorities
- company interests

Score: +1

---

**Ethical Perspective**

Examples:

- moral responsibility
- ethical duty
- doing the right thing

Score: +1

---

**Legal Perspective**

Examples:

- legal obligation
- compliance
- lawful conduct

Score: +1

---

**Regulatory Perspective**

Examples:

- regulators
- oversight agencies
- industry standards

Score: +1

---

**Technical Perspective**

Examples:

- engineering failure
- product design issue
- technical defect

Score: +1

---

**Strategic Perspective**

Examples:

- long-term positioning
- competitive strategy
- market strategy

Score: +1

---

**Social Perspective**

Examples:

- public trust
- societal impact

Score: +1

---

**Organizational Perspective**

Examples:

- communication failure
- siloed departments
- escalation problems

Score: +1

---

**Non-Scorable Responses**

The following do NOT receive points:

**Actions**

- recall the product
- offer compensation
- issue refunds
- public confession

- do nothing

---

**Procedures**

- investigate further
- gather more evidence
- contact regulators

---

**Outcomes**

- lose money
- lose reputation

These are consequences, not perspectives.

---

**Appendix B**

Participant Instructions

- AI-Prohibited Condition

Please complete this task without using AI tools, search engines, or assistance from other people.

Before writing your final response, take a few minutes to consider the situation from multiple perspectives.

As you think about the dilemma, try to consider:

- Different stakeholders
- Possible counterarguments
- Potential risks
- Alternative perspectives

There are no right or wrong answers.

We are interested in your reasoning process.

- AI-Optional Condition

You may use a generative AI tool (such as ChatGPT, Claude, or Gemini) while completing this task.

Before writing your final response, take a few minutes to consider the situation from multiple perspectives.

As you think about the dilemma, try to consider:

- Different stakeholders
- Possible counterarguments
- Potential risks
- Alternative perspectives

Please do not simply copy and paste AI responses.
Write your final response in your own words.
There are no right or wrong answers.
We are interested in your reasoning process.

- AI-Mandatory Condition

You must use a generative AI tool (such as ChatGPT, Claude, or Gemini) while completing this task.

Please use the AI tool before writing your final response.

Before writing your final response, take a few minutes to consider the situation from multiple perspectives.

As you think about the dilemma, try to consider:

- Different stakeholders
- Possible counterarguments
- Potential risks
- Alternative perspectives

Please do not simply copy and paste AI responses.
Write your final response in your own words.
There are no right or wrong answers.
We are interested in your reasoning process.

**Appendix C**

Ethical Dilemma Scenario

You are a middle manager at a medium-sized company.

While reviewing internal documents, you discover that the company is aware of a product issue but has chosen not to fully disclose it to customers.

The issue has not been officially ruled illegal, but some customers may have already experienced harm.

You raised the issue internally, but the company does not want to disclose it because of concerns about:

- Damage to the company's reputation and stock price
- Large-scale refunds and lawsuits
- Possible restructuring or job losses
- The claim that competitors are handling similar issues in the same way

However, continuing to withhold the information may expose customers to further harm and could damage public trust in the long term.

How do you think the company should respond?

Please explain your reasoning.

**Appendix D**

Question Set

Q1. What should be done in this situation?
Please explain your reasoning.

Q2. Which stakeholders should be considered in this situation?

Q3. What arguments could be made against your position?

Q4. What risks should be considered?

Q5. Are there alternative ways to view this problem?

How long did this task take to complete?

**Additional Question (AI Optional)**

How did you use AI during this task?

Please briefly describe how you used AI while completing this task. (Optional)

**Additional Question (AI Mandatory)**

1. How did you use AI during this task?

Please briefly describe how you used the AI tool before writing your final response.

2. Please paste the first AI conversation you had while completing this task.

This should include:

• Your first message to the AI
• The AI's first response

3. If you would like, please paste all or part of the AI conversation you used while completing this task.

You may paste only the part of the conversation that you found most useful, or any portion that you are comfortable sharing.

If the conversation contains personal, sensitive, or unrelated information, you may remove those portions before submitting. (Optional)